\documentclass[final,5p,twocolumn]{elsarticle}

\usepackage{graphicx,bm,amsmath,amssymb}

\journal{and accepted in Computer Physics Communications}

\begin{document}

\begin{frontmatter}

\title{\textbf{Improved parallelization techniques for the density matrix renormalization group}}

\author{Juli\'{a}n~Rinc\'{o}n\corref{cor1}}
\cortext[cor1]{Corresponding author. Tel.: $+54$ $2944$ $445378$}
\ead{julian.jimenez@cab.cnea.gov.ar}

\author{D.~J.~Garc\'{\i}a, and K.~Hallberg}
\address{Centro At\'{o}mico Bariloche and Instituto Balseiro, Comisi\'{o}n Nacional de Energ\'{\i}a At\'{o}mica, CONICET, 8400 Bariloche, Argentina}

\date{\today}

\begin{abstract}
A distributed-memory parallelization strategy for the density matrix renormalization group is proposed for cases where correlation functions are required. This new strategy has substantial improvements with respect to previous works. A scalability analysis shows an overall serial fraction of $9.4\%$ and an efficiency of around $60\%$ considering up to eight nodes. Sources of possible parallel slowdown are pointed out and solutions to circumvent these issues are brought forward in order to achieve a better performance.
\end{abstract}

\begin{keyword}
density-matrix renormalization \sep distributed programming \sep MPI.

\PACS 71.10.Fd \sep 75.40.Mg \sep 71.27.+a \sep 78.67.Hc

\end{keyword}

\end{frontmatter}


\section{Introduction}
The impact of numerical methods in the study of phenomena which are hardly understood by means of analytical machinery has been decisive. Hence, the current algorithms ought to be constantly assessed regarding the emergence of new concepts and the increasing computing technology. Nowadays one of the most successful algorithms dealing with one-dimensional interacting systems is the so-called Density Matrix Renormalization Group (DMRG)~\cite{white}. Although this method is not, strictly speaking, a renormalization procedure, the key idea is the decimation of the Hilbert space by appealing to the concept of the reduced density matrix. This fundamental concept has permitted implementing the DMRG to an extensive variety of systems and physical problems such as small grain physics, classical 2D systems, nuclear physics, quantum information, quantum chemistry, bosonic and fermionic degrees of freedom, and spin systems, together with finite temperature and non-equilibrium problems~\cite{scholl,bookdmrg}.

In most of the interesting physical situations, one has to deal with very large systems in order to prevent, for instance, finite-size effects. This fact leads unavoidably to exhaust single-machine resources. Additionally, as the dimension of the problem increases, the computational costs become more demanding. Bearing this in mind, it seems natural to request for a distributed kind of calculation. Earlier proposals consisted on shared-memory approaches~\cite{jeckelmann} for the DMRG: this method was based on the multithreaded API (Application Programming Interface), namely, OpenMP~\cite{omp}. Distributed-memory versions of DMRG have been recently proposed in several contexts~\cite{chan,kura,yama,alva}. For DMRG calculations in quantum chemistry very powerful parallel algorithms have been proposed with two basic approaches: $(i)$ the clever distribution of the local, doubly, and triply contracted orbital operators with an almost linear speedup~\cite{chan}, or $(ii)$ the dynamical scheduling of the sub-blocks of the orbital operators labeled by their corresponding quantum numbers~\cite{kura}. Concerning strongly correlated systems, there have been a few solutions to handle two-dimensional geometries by coding a parallelization that converts the superblock vectors into distributed matrices~\cite{yama}, or a generic version of a one-dimensional DMRG including a parallelization over symmetry-related matrix blocks~\cite{alva}.

The main idea behind these methods was to parallelize the central operation of a ground-state DMRG simulation: the matrix-vector multiplication in the diagonalization of the superblock Hamiltonian. However, the scheme does not take into account calculations of measurements such as expectation values, multiple-point correlation functions, and structure factors, for which the most time-consuming part of the algorithm is the huge amount of matrix-matrix multiplications (i.e.\ density-matrix rotations) of the operators one is interested in. In addition, the shared-memory scheme would already show scalability problems in a large-scale computation including the calculation of such physical quantities.

In this work, in addition to recoding the ground-state DMRG in the well-known passing message standard MPI~\cite{mpi} (henceforth \emph{regular parallelization}), we propose an im\-pro\-ved strategy that takes into account the heavy rotations associated to the calculation of the correlation functions; this policy is also implemented in MPI allowing us to perform genuine high-performance simulations~\cite{chan}. Two approaches to deal with these rotations are proposed. The first strategy is based on a pool of tasks in which there is a master node distributing queues to the rest of the slaves. The second application performs a block-fashion single distribution considering all nodes with an equal amount of work, hereafter the \emph{uniform-matrix distribution} (UMD) strategy. The latter is easier to implement and more efficient than the former. We obtain similar results for the speedup and performance to previously reported ground-state DMRG simulations with OpenMP. The chosen benchmark was the one-dimensional Hubbard model~\cite{hubb}.

In the forthcoming sections the DMRG algorithm will be briefly described, then the usual and new parallelization strategies will be presented and speedup/performance results are analyzed. Thereupon, an application test on the Hubbard model is done to estimate the runtime improvement due to the parallelization ideas of the previous sections, and we finally summarize significant concepts.

\section{The DMRG algorithm\label{sec:dmrg}}
This variational, non-perturbative and highly accurate method~\cite{scholl} was developed as an attempt to solve the low-lying energy properties of many-body models that techniques such as exact and Lanczos diagonalization~\cite{Gagliano}, numerical renormalization group (NRG)~\cite{Wilson} or other analytical tools could not be able to deal with; moreover this method does not have the sign problem that emerges in Monte Carlo techniques~\cite{Dagotto}. It can be considered as an improved version of Wilson's NRG for which the states kept during the decimation procedure are no longer selected regarding their energy but instead, they are chosen by means of the density matrix, which naturally gives the most relevant states to be kept (with respect to, e.g.\ the lowest-lying eigenstate of the whole system).

\begin{figure}
\centering\includegraphics*[width=8.5cm]{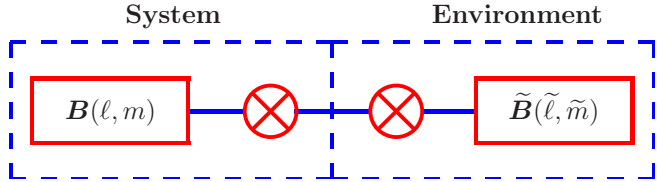}
\caption{DMRG block configuration. The superblock is formed with two blocks ($\bm B$ and $\bm{\widetilde B}$) and two (exact) sites. Added sites $\bm a$ and $\bm {\widetilde a}$ are shown as circles. Dashed lines represent \emph{system} and \emph{environment} blocks. The tilde on the right block means that no reflection symmetry has been assumed.}
\label{fig1}
\end{figure}

The standard configuration used in the DMRG algorithm is shown in Fig.~\ref{fig1}. We assume the following notation: $\bm B(\ell,m)$ a block composed of $\ell$ sites with a Hilbert space of dimension $m$ and $\bm a(\ell',m')$ a small added block (usually a single site, e.g.,\ for the Hubbard model case: $\ell'=1$ and $m'=4$). Therefore, the \emph{superblock} is formed by the union of two blocks and two sites as shown in Fig.~\ref{fig1}. This superblock is built up of two main parts: the \emph{system} and the \emph{environment} composed by a block-site each. $\bm B(\ell,m)$ is a vector space with a completeness relation close to but not equal to $\mathbf 1$ due to the decimation process.\footnote{~For the reader not familiar with DMRG, blocks and sites can be thought of as vector spaces on which there are certain conditions for well-defined states and operators.} On the contrary, the subspace $\bm a$ is always complete.

The main goal is typically the lowest-energy (ground) state of the superblock Hamiltonian $H$ which can be written as
\begin{equation}
|\psi_0\rangle = \sum_i\sum_j \psi_{0,ij}|i\rangle\otimes|j\rangle ,
\end{equation}
where $\{|i\rangle\}$ and $\{|j\rangle\}$ stand for the orthonormal basis for the \emph{system} and the \emph{environment} respectively and $\psi_{0,ij}=\langle i\otimes j | \psi_0 \rangle$. A truncation procedure should be now established in order to get manageable Hilbert spaces. To this end, DMRG resorts to the reduced density matrix of the \emph{system}:
\begin{equation}
\rho_{ii'} = \sum_j \psi_{0,ij}\,\psi_{0,i'j}^{*}.
\label{rho}
\end {equation}
This matrix possesses non-negative eigenvalues $w_\alpha$ with eigenvectors $|w_\alpha\rangle$ $\left(\rho\, |w_{\alpha}\rangle = w_{\alpha}\,|w_{\alpha}\rangle\right)$. It can be shown~\cite{bookdmrg} that these eigenvalues are proportional to the probability of the \emph{system} being in the state $|w_\alpha\rangle$. Selecting the corresponding eigenstates which have the largest probabilities $w_\alpha$, we can set a cutoff such that we have a very efficient decimation formula. This error source can be quantitative described by defining the \emph{truncation error}
\begin{equation}
\epsilon_\rho = 1-\sum_\alpha^m w_\alpha ,
\label{error}
\end {equation}
where $m$ is the cutoff, a truncation number selected often by hand. It can be shown~\cite{scholl,bookdmrg} that the error in the ground state goes as $\| |\overline\psi_0\rangle - |\psi_0\rangle \|^2=\epsilon_\rho$ where $|\overline\psi_0\rangle$ is the DMRG approximation to the exact ground state. A similar bound can be found for the expectation values. It is also shown that the energies obtained with DMRG will be upper bounds on the exact eigenvalues. From Eq.~(\ref{error}) it is evident that the more states are kept the higher the accuracy of the calculated energies and observables will be. Another (generally smaller) source of error in $|\psi_0\rangle$ is due to the iterative method used to diagonalize the superblock Hamiltonian. As a consequence of the Hilbert space truncation there is an \emph{environmental error} which has to do with the fact that the bath coupled to the \emph{system} is not exact. The environmental error can be reduced by implementing the so-called finite system algorithm.

The arrangement shown in Fig.~\ref{fig1} is usually used in two ways: on one hand, the \emph{infinite system algorithm} in which the superblock size is grown by adding two new sites in the middle of the chain at each iteration step. And on the other hand, the \emph{finite system algorithm} is designed to calculate highly accurate properties of the superblock at a given lattice length. It consists on moving back and forward (sweeping) the division between \emph{system} and \emph{environment} (it can be thought of as a thermalization of the \emph{system} and \emph{environment} blocks).

All these steps can be summarized in the following way:
\begin{enumerate}
\item Start with left and right blocks as exact single sites.
\item Diagonalize the superblock Hamiltonian $H$ defined on $[ \bm{B}(\ell ,m)\; \bm{a}\; \bm{\widetilde a}\; \bm{\widetilde B}(\widetilde\ell,\widetilde m) ]$ to obtain $|\psi_0\rangle$.
\item Build up all of the block operators related to $H$ and measurements defined on $\left[\bm{B}\, \bm{a}\right] \doteq \left[\bm{B}\oplus \bm{a}\right]$.
\item Define and diagonalize $\rho$ in the \emph{system}. Find the rotation matrix $\mathcal{R} = (|w_1\rangle |w_2\rangle \cdots |w_m\rangle)^T$ formed from the $m$ largest eigenvalues $w_{\alpha}$ of $\rho$.
\item Perform the decimation and rotation step $\left[\bm{B}\right]\longleftarrow \mathcal R\left[\bm{B}\, \bm{a}\right]\mathcal R^+$ for the operators defined in step 3. \\Go to step 2.
\end{enumerate}

When the desired system size has been achieved, measurements of the relevant quantities such as structure factors, spin and charge gaps, binding energies, etc.\ can be performed.

Since our main concern is the computation of $n$-point correlation functions for several operators $Z$, we have to provide a form for such matrices. This type of simulation can be included in the standard algorithm just managing those $Z$ operators in the same way as the superblock Hamiltonian operators are handled, that is, by doing the transformations of blocking $Z_{\left[\bm{B}\, \bm{a}\right]}\longleftarrow Z_{\left[\bm{B}\oplus \bm{a}\right]}$ and then the rotation and the decimation step $Z_{\left[\bm B\right]}\longleftarrow \mathcal RZ_{\left[\bm{B}\, \bm{a}\right]}\mathcal R^+$. All of the operators $Z$ are managed as block matrices instead of as block-site matrices reducing the consumed computational resources and saving time on I/O operations.

\subsection{Benchmark}
We have tested the parallel algorithm with a simulation of the one-dimensional quarter-filled Hubbard model~\cite{hubb}. The Hamiltonian of the model reads: 
\begin{equation}
H = -t\sum_{i,\sigma} \left( c^+_{i+1\sigma}c_{i\sigma} + c^+_{i\sigma}c_{i+1\sigma} \right) + \frac{U}{2}\sum_{i,\sigma}c^+_{i\sigma}c_{i\sigma}c^+_{i\bar\sigma}c_{i\bar\sigma} 
,
\label{HUBB}
\end{equation}
where $c_{i\sigma}$ ($c^+_{i\sigma}$) denotes an electron annihilation (creation) operator on site $i$ with spin $\sigma=(\uparrow,\downarrow)$. Here, $c_{i\sigma}$ is an $m\times m$ matrix. Regarding storage effects, $\sigma$ implies two different matrices for $c_{i\sigma}$ for each site $i$. $t$ and $U$ are parameters standing for electron hopping and on-site electron repulsion respectively.

The charge $N(q)$ and spin $S^z(q)$ structure factors 
\begin{equation}
\begin{split}
N(q) &= \frac{1}{L}\sum_{k,j} e^{iq(k-j)}\langle (n_k - n)(n_j - n) \rangle \\
S^z(q) &= \frac{1}{L}\sum_{k,j} e^{iq(k-j)}\langle S^z_kS^z_j \rangle 
\end{split}
\end{equation}
were calculated, the number operator is $n_{i\sigma}=c^+_{i\sigma}c_{i\sigma}$, $n_i = n_{i\uparrow} + n_{i\downarrow}$, and $n$ is the charge expectation value. As it can be seen, obtaining these two quantities requires the calculation of the expectation values $\langle n_i\rangle$ and all of the charge-charge $\langle n_in_j\rangle$ and spin-spin $\langle S^z_iS^z_j\rangle$ correlation functions.

\section{Parallelization}
There are two main architecture paradigms in parallel computing: systems with a single address space called \emph{shared-memory systems} allowing multiple processors to access the same memory location (data) and \emph{distributed-memory systems} in which each processor has its own address space and therefore its own data structure. Both paradigms can be successfully applied to the DMRG me\-thod~\cite{jeckelmann,chan}. Earlier distribution strategies worked well on a shared-memory system methodology; nevertheless, this type of architecture eludes a massively parallel approach. Consequently, a distributed-memory policy should be developed in order to get a coarse-grain scheme reaching larger lengths and more states per block using modest computational resources. Here, in addition of putting forward a new parallelization scheme, we have changed the shared-memory (OpenMP) approach to a standard message passing API (MPI)~\cite{mpi}.

As we will show below very similar results are obtained to the OpenMP case with the possibility of improving scalability properties. This distributed approach has the advantage of avoiding collisions (present on MP algorithms) at the presumable cost of using more resources and larger communications. The calculations presented in this work were performed using a cluster with Intel$^\circledR$ Xeon $2.50$~GHz CPU cores (with a memory of 1~GB per node) arranged either as a double quad-core system or as single cores in a star topology network with a nominal bandwidth of $900$~Mb/s.

Let us now briefly summarize the analytical apparatus needed to study the speed of a high-performance realization~\cite{book}. The \emph{speedup} $S_p$ indicates how much faster a parallel code on a $p$-node process is with respect to the sequential analogue. $S_p$ is explicitly defined as the fraction
\begin{equation}
S_p = \frac{T_1}{T_p},
\end{equation}
where $T_1$ and $T_p$ are the wall-clock times of the simulation with $1$ and $p$ processors respectively. The ideal speedup should scale linearly with $p$, that is, $S_p^{\mathrm{ideal}} = p$. Another quantity of interest which illustrates how much the algorithm is exploiting a single processor is the \emph{efficiency} which reads 
\begin{equation}
E_p = \frac{S_p}{p}.
\end{equation}
In the simplest model, the sequential time of a program (normalized to $1$) can be split into a \emph{serial fraction} $\Sigma$ and a \emph{parallel fraction} $1-\Sigma$. With a finite number of nodes $p$, the parallel fraction gets reduced by $(1-\Sigma)/p$; based on these considerations we obtain \emph{Amdahl's law}~\cite{amdahl} for the relative speedup 
\begin{equation}
 S_A(p) = \left(\Sigma + \frac{1 - \Sigma}{p}\right)^{-1},
\label{amh}
\end{equation}
thus, the maximum speedup achievable (i.e.\ with $p\rightarrow\infty$) would be $S_A \rightarrow 1/\Sigma$. This amount gives us a rough idea of the expected efficiency in a distributed implementation.\footnote{~This fixed-sized problem law neglects important effects such as overhead, cache effects, network latency, etc.}

\subsection{Regular Parallelization: Ground-state DMRG}
It is well known that the most time-consuming part in the ground state DMRG is obtaining the lowest eigenvalue of the superblock Hamiltonian $H$ by means of an iterative procedure (such as Lanczos~\cite{Gagliano} or Davidson~\cite{david} algorithms). Since $H$ is actually a sum of terms involving left (\emph{system} formed by $\bm B\oplus \bm a$) and right (\emph{environment} formed by $\bm{\widetilde a}\oplus\bm{\widetilde B}$) matrix products, we can readily write 
\begin{equation}
H = \sum_\lambda O^\lambda_{\bm B} \otimes O^\lambda_{\bm a} \otimes O^\lambda_{\bm{\widetilde a}} \otimes O^\lambda_{\bm{\widetilde B}},
\end{equation}
where $O^\lambda_{\bm X}$ represents a generic operator defined on any of the blocks ($\bm X = \bm B, \bm a, \bm{\widetilde a}, \bm{\widetilde B}$) and $\lambda$ corresponds to each of the terms in Eq.~(\ref{HUBB}). Typical terms are for instance, the hopping term between the left block and left site: $c^+_{\bm B}c_{\bm a}\equiv c^+_{\bm B}\otimes c_{\bm a}\otimes I_{\bm{\widetilde a}}\otimes I_{\bm{\widetilde B}}$ or the right block Hamiltonian: $H_{\bm{\widetilde B}}\equiv I_{\bm B}\otimes I_{\bm a}\otimes I_{\bm{\widetilde a}}\otimes H_{\bm{\widetilde B}}$ which should contain all of the $H$ terms for the sites belonging to $\bm{\widetilde B}$. $I_{\bm X}$ stands for the identity on the space $\bm X$.

If the implementation incorporates symmetries, such as particle number or total magnetization, then $H$ takes the form
\begin{equation}
 H = \sum_\lambda\sum_{\theta} O^\lambda_{\bm B}(\theta^{\bm B}) \otimes O^\lambda_{\bm a}(\theta^{\bm a}) \otimes O^\lambda_{\bm{\widetilde a}}(\theta^{\bm{\widetilde a}}) \otimes O^\lambda_{\bm{\widetilde B}}(\theta^{\bm{\widetilde B}}),
\label{hache}
\end{equation}
explicitly showing that the operators $O^\lambda_{\bm X}(\theta^{\bm X})$ are labeled by their quantum numbers. The value $\theta^{\bm X}$ is a symmetry index of the $\bm X$ block, and $\theta$ is an index running over the superblock basis formed by the configurations with the quantum number $\theta^{\bm B} + \theta^{\bm a} + \theta^{\bm{\widetilde a}} + \theta^{\bm{\widetilde B}}$ fixed. Using symmetries helps to minimize the size of nested loops. Usually $H$ is a very large matrix (e.g.\ with dimension $\mathcal M\sim 10^4-10^6$), thus it is never explicitly constructed but rather consists of multiplication rules. This means that given a vector $|b\rangle$ we get the $H$-multiplied result $H|b\rangle$.

\begin{figure}
\centering\includegraphics*[width=8cm]{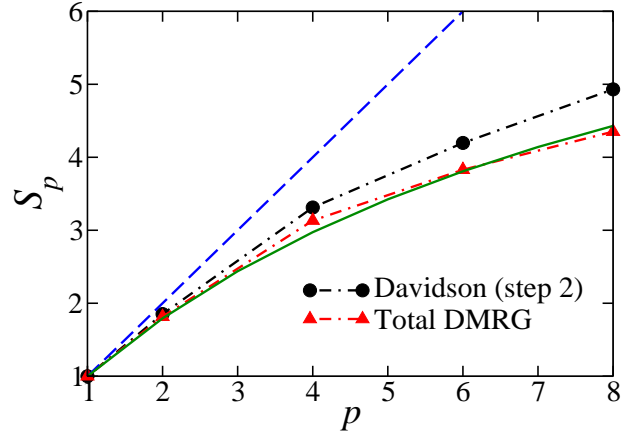}
\caption{Speedup scaling of a DMRG calculation for the ground state. Circles correspond to the Davidson algorithm (step 2 in section~\ref{sec:dmrg}) performance for the $\theta$ index distribution case $\left(\Sigma = 8.6(2)\%\right)$. Triangles correspond to the total DMRG calculation of the ground state $\left(\Sigma = 11.5(3)\%\right)$ with the corresponding Amdahl's law (full line). Ideal scaling is included for comparison (dashed line).}
\label{velgs}
\end{figure}

We shall now get into the aspects of the parallelization idea. There is a basic tactic without handling the matrix-vector multiplication which would be that of distributing only the Hamiltonian terms mentioned above, that is, the $\lambda$ index in Eq.~(\ref{hache}). Explicitly, one node will deal with $H_{\bm B}$, another node will address the $c^+_{\bm B}c_{\bm a}$ term, and so on. However, this plan is prone to poor scalability showing parallel slowdown already for $6$ nodes with a speedup of only $1.5$. This slowdown is perhaps due to load imbalance since not all of the Hamiltonian terms involve the same number of operations. The site-site interaction consists only of a few logical rules, but terms such as block-site or site-block have to iterate over tensor products. Even when we compare these last two terms there is also an imbalance because of roaming over fast and slow matrix indices. 

A more efficient option consists of the distribution over the central ($\theta = 1,\dots, \mathcal M$) loop of the matrix-vector multiplication on the diagonalization algorithm (Davidson in our case). Each task will apply the full $H$ to $\lfloor \mathcal M/p\rfloor$ states and the first $\bmod(\mathcal M,p)$ tasks will handle an extra state.\footnote{~$\lfloor x/y \rfloor$ meaning the integer division and $\bmod (x,y)$ stands for the modulo operation with $x$ and $y$ real numbers.} We do not distribute the sub-blocks of the relevant operators labeled by
their quantum numbers because of their dissimilar dimensions. With this strategy, we get values of speedup of $3.5$ in an $8$-node process with a serial fraction of $11.5\%$. To achieve an even faster realization when distributing over the $\theta$ index, one should also share out all of the linear algebra (\verb+daxpy+, \verb+ddot+, \verb+dscal+, and \verb+dcopy+) operations in the Davidson algorithm. These operations include orthonormalizations, inner products and the normalizations of the vectors added to the Davidson basis expanding the ground state $|\psi_0\rangle$. In doing so, we have now moved up the speedup to $4.9$ on 8 nodes ($\Sigma=8.6\%$). The scalability properties of the distributed version of the DMRG calculation for $|\psi_0\rangle$ are shown in Fig.~\ref{velgs}. The load imbalance in this case goes as $p/\mathcal M$ which is negligible for actual DMRG simulations.

The performance properties of Davidson parallelization are strongly affected by the reduction operations of the matrix-vector multiplication, hence the better the implementation of these the better the speedup will be. This leading behavior could be diminished by ordering the superblock basis properly. This way, all of the reduction calls of order $\mathcal M$ are optimized by calls of order $\mathcal M/p$ or less. To show this, we have used a test block-diagonal matrix that does not require any reduction calls at all in the application of $H$. By doing this, we have obtained a serial fraction of $\Sigma = 0.87(2)\%$ (down to $20$ processors) on the Davidson scheme, whereas when we consider the Hubbard Hamiltonian, we get a serial fraction of $8.6\%$ as a result.

The most simple distribution one can think of was implemented in the rotation (decimation) of the operators relevant to $H$ (such as $c_\uparrow$, $c_\downarrow$ for the Hubbard model case), that is, a row-distributed matrix-matrix multiplication. The final result is a serial fraction of $25\%$ for this section of the algorithm. The reader should remember that Amdahl's law is a very simplistic proposal on the performance of a parallelized algorithm; serial fractions allow us to easily understand the results and what to expect of a distributed version of the serial code. 

In order to better understand the performance obtained, we now make a comparison between our MPI implementation of the 1D Hubbard model and the shared-memory (OpenMP) version of the 2D Hubbard model~\cite{jeckelmann}. The who\-le DMRG performance of the MPI implementation shows a better behavior than in the shared-memory version ($\Sigma=11.5\%$ compared to $\Sigma=16\%$~\cite{jeckelmann}) in spite of the fact that the Davidson algorithm results are not as good as previous ones ($\Sigma=8.6\%$ compared to $\Sigma=6.5\%$~\cite{jeckelmann}). This improved behavior could be related to the additional parallelization of the linear algebra operations mentioned above, added to the absence of collisions (and despite message passing) on the MPI algorithm or better communications originated on newer hardware improvements. Even though this comparison is not strictly valid because we are dealing with different geometries (1D versus 2D~\cite{jeckelmann} Hubbard models), we must remark that our case is the worst case scenario. In 1D we have fewer Hamiltonian terms, meaning fewer independent processor operations in comparable Hilbert spaces with a similar amount of communications. This would suggest that for a more complex Hamiltonian (e.g.\ including longer range hoppings or different geometries such as 2D) our result for the serial fraction will be even smaller.

\begin{table}[b]
\caption{\label{tabla1}Relative runtimes and serial fractions percentages at different steps of the algorithm. The unparallelized time (item $d$) is mainly consumed in building the \emph{system} (or the \emph{environment}), the density matrix and getting its spectrum.}
\renewcommand{\tabcolsep}{0.35cm}
\begin{tabular}{lcc}
\hline\hline
Step & Time~$(\%)$ & $\Sigma~(\%)$ \\
\hline
$a.\quad$ Davidson algorithm & 19.7 & 9.4(1) \\
$b.\quad$ $Z_i$ and $Z_iZ_j$ rotations & 72.3 & 8.1(3)\\
$c.\quad$ $H$ operators rotations & 0.1 & 25(2)\\
$d.\quad$ Unparallelized sections & 0.5 & 100(0)\\
$e.\quad$ Measurements & 7.4 & 7.6(7)\\
$f.\quad$ Total calculation & 100 & 9.4(1)\\
\hline\hline
\end{tabular}
\end{table}

\subsection{Novel Strategy: Correlation operators}
If $n$-point correlations are required, the former distribution setup turns out to be insufficient because the ground state determination is not the longest time-consu\-ming part anymore and is overtaken by the operator decimation and rotation (see Table~\ref{tabla1}). Therefore a new approach is mandatory to deal with that issue. The new strategy should take into account that the most time-expensive part is in this case the double matrix operation of the corresponding operators $Z_i$ and $Z_iZ_j$ (e.g.\ for $\mathcal R Z_i \mathcal R^+$: $Z_i\mathcal R^+$ and then $\mathcal R(Z_i\mathcal R^+)$). Typical correlation functions are the one-point and two-point functions~\cite{scholl}, namely,
\begin{equation}
\begin{split}
\langle Z_i\rangle &= \langle\psi_0| Z_i|\psi_0\rangle,\\
\langle Z_iZ_j\rangle &= \langle\psi_0| Z_iZ_j|\psi_0\rangle 
\end{split}
\end{equation}
with $i,j=1,\dots ,L$ and $L$ the length of the superblock chain. The number of (stored) matrices to be rotated (see section~\ref{sec:dmrg}, last step) at a given length calculation is $\mathcal L = \ell(\ell + 3)/2$ ($\ell$ matrices coming from single-site operators $Z_i$ and $\ell(\ell + 1)/2$ coming from two-point correlation functions $Z_iZ_j$ with $i<j$), with $\ell$ being the number of sites of the \emph{system} or \emph{environment} according to forward or backward sweeping. The correlations between the $\bm B$ and $\bm{\widetilde B}$ blocks were calculated as a product of single-site operators in each block. The specific tasks involved in step 5 (see section~\ref{sec:dmrg}) are: ($i$) the reading of the current matrix $Z_i$ from storage, ($ii$) the blocking step $Z_{\left[\bm{B}\, \bm{a}\right]}\longleftarrow Z_{\left[\bm{B}\oplus \bm{a}\right]}$, ($iii$) the two matrix-matrix products with the rotation matrix $\mathcal R$, and ($iv$) the corresponding saving of the new matrix $Z_i^{\mathrm{new}} = \mathcal RZ_i\mathcal R^+$.

\begin{figure}
\centering\includegraphics*[width=8cm]{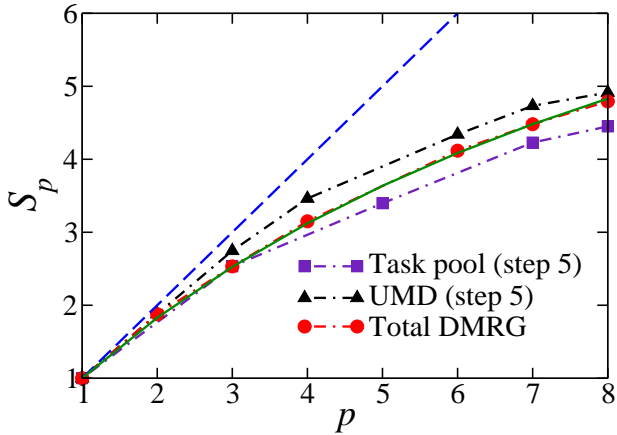}
\caption{Pool of tasks (squares) and UMD parallelization (triangles) performance for step 5 in section~\ref{sec:dmrg}. The values of the serial fractions were $\Sigma = 11.1(2)\%$ and $\Sigma = 8.1(3)\%$ respectively. The speedup factor corresponding to the total DMRG calculation using the UMD technique for the calculation of the correlation functions (circles) was $\Sigma = 9.4(1)\%$. Its corresponding Amdahl's law is also included (full line) and the ideal scaling is shown for comparison (dashed line).}
\label{velpo}
\end{figure}

\begin{figure}
\centering\includegraphics*[width=8.5cm]{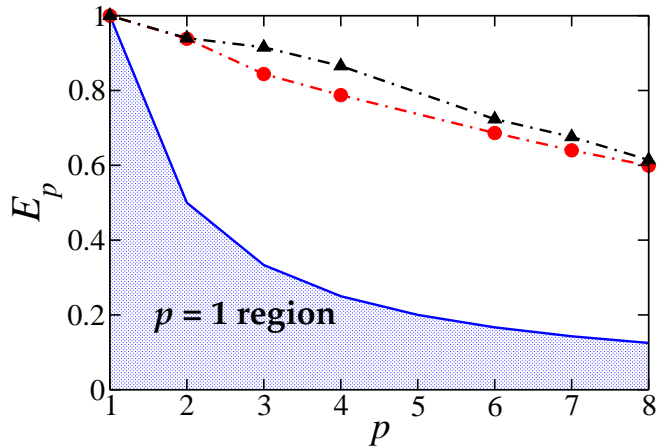}
\caption{Efficiency plot for the UMD case shown in Fig.~\ref{velpo} with the same symbol convention. The shaded region corresponds to the cases where no speedup is gained compared to the $p=1$ case.}
\label{eficiencia}
\end{figure}

We shall show below two ways to address this issue: a pool of tasks~\cite{pool} and what we have called a \emph{uniform-matrix distribution} (UMD) parallelization. In this latter strategy every node has almost the same load (see below) without a master node. The UMD parallelization seems to have a better output because it has fewer communications (only at the very beginning of the subroutine) and takes more advantage of the nodes available during the calculation (see below). The pool of tasks is a more elegant and common solution but in practice, a slower option. The speedup results for these two parallelized DMRG calculations of correlation functions are shown in Fig.~\ref{velpo}. The efficiency for the UMD case is shown in Fig.~\ref{eficiencia}.

In the pool of tasks paradigm~\cite{pool}, the data to be processed (the $Z$ matrices) are divided into small units with similar structures called \emph{tasks}. All of these tasks form the so-called \emph{task pool}. One node, the master process, manages this large amount of tasks, always sending to idle workers more work to do until all of the tasks have been executed (empty pool). This model is effective in situations where the available nodes have very different technical specifications, because the least loaded or more powerful hosts do more of the work and all of the hosts stay busy most of the runtime. The serial fraction obtained in this implementation was about $11.1\%$ (see Fig.~\ref{velpo}). The optimal result depends on the number of tasks in which the whole job is divided. If this number is too small, parallel slowdown will already appear. In addition, the greater the number of tasks the bigger the amount of communications will be.

Let us now explain the UMD technique. This distribution proves to be easier to code and more efficient than the pool of tasks. The key idea is to keep all of the processors on the same working settings so we can take full advantage of the accessible hardware. The distribution is performed in terms of blocks of contiguous local and non-local operators. If the number of processors is $p$ then each processor stores $\left\lfloor\mathcal L /p\right\rfloor$ operators, except maybe the first $\bmod(\mathcal L,p)$ ones that will store $\left\lfloor\mathcal L /p\right\rfloor + 1$ matrices. Load imbalance in this case goes as $p/\mathcal L$, which is imperceptible for larger lattice lengths, i.e.\ larger $\mathcal L$. The serial fraction has now been improved to $\Sigma = 8.1\%$ (in the double quad-core system) as shown in Fig.~\ref{velpo}.

There are many more communications in the pool of tasks compared to the UMD case. These communications are related to petitions coming from the workers involving statuses such as: ``task done'' and ``ready to work''; and the complementary messages sent by the master node with the proper information about the task to be made. On the contrary, the UMD settings just need very few communications that keep track of the set of operators to be handled by each node. This message passing should be posted at the beginning of the corresponding iteration.

In both parallelization policies, if a given node demands a specific set of matrices that is not currently in local storage, an implemented queue manager handles this type of requests by sending the matching operator. This is done by means of a book-keeping of the matrices and its current owners throughout the entire cycle. Hence, when all of the desired matrices have been shipped, a new-owner message should be broadcasted to the rest of the active processors. The rotation matrix $\mathcal R$ is replicated along all of the nodes. This procedure allows each processor to save runtime by storing the new operators locally. For instance, if at some point through the simulation a processor, say, number $1$ requests an operator that in an earlier step was assigned to processor, say, number $2$, the queue handler transfers the required matrix from processor $2$ to the corresponding node making an update of the owner matrix-bookkeeping. This procedure does not affect the task being performed by processor $2$ avoiding synchronization delays. For the UMD case we have found a serial fraction of $\Sigma = 9.4\%$ in the star topology network. 

The origin of the serial fraction of the presented parallelization schemes is perhaps due to the following factors: processes contending available cache space, racing conditions linked to the storage of the corresponding matrices, or the transfer of the requested data between processes. In order to reduce the total serial fraction of the whole process attention should be paid to the rotations of the operators (item $b$ in Table~\ref{tabla1}), the Davidson algorithm (item $a$), and the unparallelized sections (item $d$). The measurements are discussed below. As for item $b$, the most time-expensive of all of the four steps at this point (addressed at the beginning of this subsection) would be consecutively: the two matrix-matrix products, the writing of the outcome to disk, the reading of the input from disk, the blocking operation and, in the star-topology case, the matrix copying among nodes. Unavoidable points are probably the I/O operations, the matrix multiplications, and the optimized blocking due to the use of symmetries. Therefore the candidate stage to be improved is the data transfer protocol (\verb+ssh-server+) for the networking case. Using a socket-type communication or a remote server will certainly enhance the achieved speedup. As for the Davidson step, in all of the strategies, one could try to reduce the few synchronization calls with the consequence of having more local operations. And finally, the total serial fraction could be reduced further if some kind of parallelization scheme is implemented in the unparallelized section of item $d$.


There is a small discrepancy between the values of the serial fractions shown in Fig.~\ref{velgs} and Table~\ref{tabla1}, item $a$ (with and without correlations) for the Davidson part. This may be due to the effect of the compilation when correlations are included. However, the values are compatible within the numerical error. Now, taking into account the Davidson diagonalization, as well as the Hamiltonian operators and the rotation of the operators to be measured, we should get a weighted average serial fraction of $\Sigma_{\textrm{total}} = 8.8\%$ as for the parallelized sections, but due to the unparallelized fraction of the code (item $d$) the final serial fraction is actually $9.4\%$. Finally, the corresponding distribution was done for the measurement part in the same way as for the distribution over the $\theta$ index in Eq.~(\ref{hache}), with the exception that the $\mathcal M$-size vector \verb+reduce+ calls have been replaced by single-data reductions associated to the partial inner products $\langle\psi_0|Z|\psi_0\rangle$. The serial fraction for this section of the algorithm was $\Sigma = 7.6\%$. This value is probably related to the reading of the $Z$ matrices from local or remote storage depending on the final $Z$-bookkeeping. It should be mentioned that this is just a minor optimization compared to the whole calculation, but it is rather straightforward to code this section of the DMRG algorithm once that of the Davidson diagonalization has been implemented.

To estimate the performance of each node as compared to communication times, we show in Fig.~\ref{eficiencia} the parallel efficiency of the whole process in the UMD case. This quantity shows a very nice behavior up to the number of nodes used. For the $p=8$ case $E_p$ is around $60\%$ meaning that each processor is actually working more than half of the total computational time. It also shows the good reliability of the parallelized algorithm suggested in this work. Parallel efficiency of a single-CPU is shown for comparison (continuous line). An improvement in the overall efficiency was observed when the number of states kept was increased $(m = 400-1000)$, as expected from a non-fixed-sized parallel problem~\cite{gusta}. For instance, for $m=1000$, $E_p$ is increased by $20\%$ for $p=8$ with an overall serial fraction of $\Sigma = 5.5\%$. It should be noticed that the more operators are measured the more effective this novel strategy will be.

Simulations of ladder-type systems have shown that the ratio of runtimes between Davidson diagonalization and the rotation of the operators is not as remarkable as in the one-dimensional case. However, for long enough systems, the time of the rotation of the operators will be a significant part of the total time justifying the implementation of the present parallelization strategies. The change of the Davidson runtime stems from the increasing number of terms of $H$ as pointed out in the previous subsection.

Lastly, in order to reproduce well-known results for the $S^z(q)$ and $N(q)$ structure factors~\cite{krho}, we have performed serial and distributed numerical simulations for a quarter-filled one-dimensional Hubbard chain of $L=128$ sites with $m=400$ states per block and an interaction parameter $U/t=8$. Two sweeps for the finite-size algorithm and open boundary conditions were imposed in the calculation. The truncation error was $\epsilon_\rho \sim 10^{-7}$. The total runtime on a $1$-node process was about $165$ hours compared to, for instance, $33$ hours on an $8$-node process.

\section{Conclusions}
We have presented an efficient parallelized version of a DMRG code devoted to the calculation of $n$-point correlation functions. Unlike previous approaches, the current strategy was implemented in a passing message context (MPI) allowing for a better performance than for the shared-memory scheme. The overall serial fraction of the whole process was about $9.4\%$ and the efficiency was around $60\%$ up to eight nodes. In spite of the fact that our parallelization scheme does not scale well to hundreds of nodes it does allow simulations not reachable by serial coding with a maximum speedup of $1/\Sigma = 10.6$ according to Amdahl's law. Causes of parallel slowdown were addressed and possible ways of decreasing the serial fraction were presented.

\section*{Acknowledgments}
J. R. would like to thank to E. Dari and E. Tapia for useful discussions and is infinitely indebted to P. Mateo for unconditional support. This work was done in the framework of projects PIP~5254 of the CONICET and PICT~2006/483 of the ANPCyT.

\end{document}